\documentclass[floatfix,prl,english, twocolumn,superscriptaddress]{revtex4-1}
\usepackage{graphicx}
\usepackage{graphics}
\usepackage{braket}
\usepackage{mathtools}
\usepackage[titletoc]{appendix}
\usepackage{appendix}
\usepackage{float}
\usepackage{bbold}

\usepackage[dvipsnames]{xcolor}
\definecolor{darkblue}{rgb}{0.0, 0.0, 0.75}

\usepackage{color}
\usepackage[colorlinks=true,
            linkcolor=darkblue,
            urlcolor=darkblue,
            citecolor=darkblue]{hyperref}

	\usepackage[normalem]{ulem} %makes underlining possible: \uline \uwave \sout
	\definecolor{mgreen}{RGB}{1,123,0}

\def \m2D{\mathrm{2D}}

\def \mexact{\mathrm{exact}}

\def \gt{\mathrm{\tilde{g}}}

\def \mcl{\mathrm{cl}}

\def \mTr{\mathrm{Tr}}

\def \v0{ {\tilde{v}_0 }}

\def \gt {\tilde{g} }

\begin{document}
%\title{Analyzing critical behavior using the open system discrete truncated Wigner approximation}
\title{Driven-dissipative criticality within the discrete truncated Wigner approximation}
\author{Vijay Pal Singh}
\affiliation{Institut f\"ur Theoretische Physik, Leibniz Universit\"at Hannover, Appelstra{\ss}e 2, 30167 Hannover, Germany}
\affiliation{Zentrum f\"ur Optische Quantentechnologien and Institut f\"ur Laserphysik, Universit\"at Hamburg, 22761 Hamburg, Germany}
\author{Hendrik Weimer}
\affiliation{Institut f\"ur Theoretische Physik, Leibniz Universit\"at Hannover, Appelstra{\ss}e 2, 30167 Hannover, Germany}
\date{\today}
%
%%%------------------------------------------ ABSTRACT ----------------------------------------------------------------------------------
%

\begin{abstract}

  We present an approach to the numerical simulation of open quantum
  many-body systems based on the semiclassical framework of the
  discrete truncated Wigner approximation. We establish a quantum jump
  formalism to integrate the quantum master equation describing the
  dynamics of the system, which we find to be exact in both the
  noninteracting limit and the limit where the system is described by
  classical rate equations. We apply our method to simulation of the paradigmatic dissipative Ising model, where we are able to capture the critical fluctuations of the system beyond the level of mean-field theory.    
\end{abstract}

\maketitle
%
%---------------------------------------------------------------Introduction: BEGIN -------------------------------------------------------
%

The identification of phase transitions and their universality classes
is one of the most important tasks in many-body physics, especially
for non-equilibrium systems where many of the conventional methods
cannot be applied. Here, we show that a large class of steady state
phase transitions arising in open quantum systems can be efficiently
simulated and analyzed using an open system variant of the discrete
truncated Wigner approximation.

Open quantum many-body systems are not only useful for the
dissipative preparation of tailored quantum many-body states
\cite{Diehl2008,Verstraete2009,Weimer2010,Krauter2011,Barreiro2011,Carr2013a,Rao2013,Morigi2015,Reiter2016,Roghani2018,Raghunandan2020,Metcalf2020}, but are also of fundamental interest, as their dynamics
can realize non-equilibrium phenomena that are not found in their
closed counterparts. Most strikingly, the steady state of an open
system can undergo phase transitions \cite{Kasprzak2006,Amo2009,Hartmann2010,Baumann2010,Nagy2010,Diehl2010a,Tomadin2011,Lee2011,Kessler2012,Honing2012,Honing2013,LeBoite2013,Horstmann2013,Torre2013,Qian2013,Lee2013,Joshi2013,Lang2015,Marino2016,Marcuzzi2016,Weimer2017,Parmee2018,Owen2018,Jamadagni2020}, where an associated
order parameter changes across the transition in a non-analytic way. A
large class of such transitions is governed by a dynamical symmetry
rendering static correlation functions to obey thermal statistics
\cite{Sieberer2013,Maghrebi2016}. Of particular interest is a
dissipative variant of the Ising model in a transverse field
\cite{Lee2011} because of its relevance for ongoing experiments with
driven-dissipative Rydberg gases \cite{Carr2013,Malossi2014}. For this
model, a first-order liquid-gas transition has been reported, which
has been predicted to end in an Ising critical point based on
mean-field calculations \cite{Marcuzzi2014}. However, since the
numerical analysis of critical open many-body systems is extremely
challenging \cite{Weimer2021}, a reliable assesment of its critical
behavior is still lacking.

In this Letter, we build upon the discrete truncated Wigner
approximation \cite{Schachenmayer2015} and introduce a variant capable
to treat open quantum systems. Our approach constitutes a wave
function Monte-Carlo method in the quantum-jump formalism
\cite{Dalibard1992,Dum1992,Molmer1993}. Crucially, our method is exact
in the non-interacting limit, which we use for benchmarking, as well
as in the fully classical limit, where coherences in the density
matrix vanish and the dynamics is governed by classical rate
equations. We then apply our method to the dissipative Ising model on
a square lattice, where we find that the transition belongs to the
two-dimensional Ising universality class. Remarkably, we obtain
critical exponents beyond their mean-field value, although the
interaction is only taken into account on a mean-field level. We
connect this surprising result to the fact that classical fluctuations
are correctly taken into account, while quantum fluctuations are
irrelevant at the transition. This scenario is characteristic for all
open quantum systems posessing the aforementioned dynamical symmetry,
hence our method can be expected to correctly describe the critical
behavior of a large class of dissipative many-body models, e.g. the dissipative XYZ model \footnote{See Supplemental Material for the derivation of the equations of motion and the jump probability, the comparison with exact results, and the OSDTWA analysis of the dissipative XYZ model.}.

\begin{figure}[t]
\includegraphics[width=1.0\linewidth]{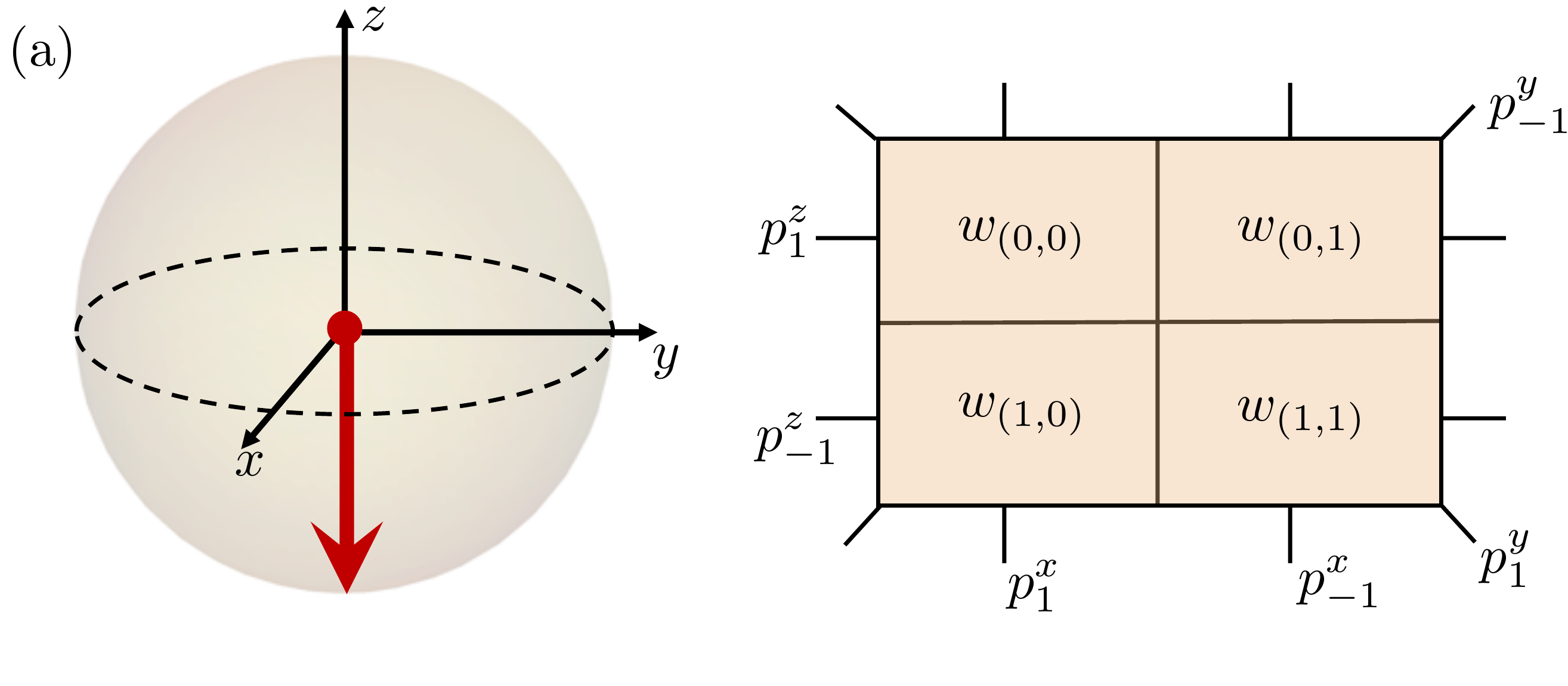}
\includegraphics[width=1.0\linewidth]{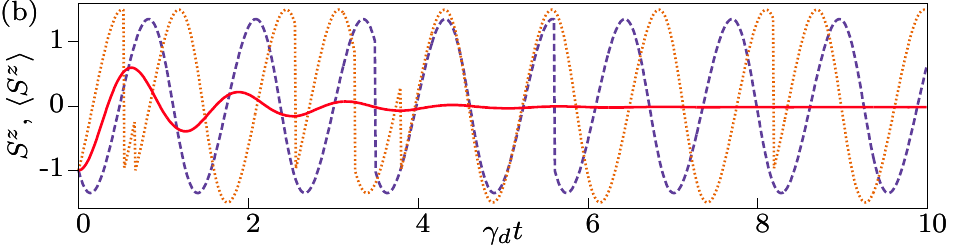}
\caption{Open-system dynamics within the discrete truncated Wigner approximation. (a) Bloch sphere representation for a spin-$1/2$ particle, where the spin points in the $-z$ direction. 
This initial state is sampled from a discrete four-point  Wigner quasiprobability distribution $w_{(p,q)}$, which are $w_{(0,0)}  = w_{(0,1)}  =0 $ and  $w_{(1,0)} = w_{(1,1)} =1/2$. 
The probability for a spin to point along the $\pm x$,  $\pm y$, and $\pm z$ directions $(p_{\pm 1}^{x,y, z})$ is given by the sum over the vertical,  diagonal, and horizontal lines, respectively \cite{Wootters1987, Schachenmayer2015}.  
(b)  Classical trajectories corresponding to two different initial configurations (dotted and dashed lines) for single spin and $g/\gamma_d = 5$. 
The averaged time evolution of $\langle S^z (t) \rangle$ over $10^{5}$ trajectories is shown as a continuous line.
 }
\label{Fig:dtwa}
\end{figure}

{\it{Open-system discrete truncated Wigner approximation (OSDTWA).}---} 
Phase-space methods, such as the truncated Wigner approximation (TWA), approximate the quantum-mechanical dynamics by  a semiclassical evolution of individual trajectories. In the TWA, which has also been employed to investigate open quantum systems \cite{Carusotto2005,Carusotto2013,Dagvadorj2015,Vicentini2018,Huber2021a}, the initial state is sampled from a continuous Wigner function \cite{Polkovnikov2010}, which is replaced by a discrete Wigner functions for systems with discrete degrees of freedom \cite{Wootters1987}. 
For a single spin-1/2 particle, we represent the discrete phase space by four phase points $\alpha = (q, p) \in \{ (0,0),(0,1), (1,0), (1, 1) \}$ \cite{Wootters1987, Schachenmayer2015, Czischek2018}. 
The corresponding phase-point operators $\hat{A}_\alpha$ are written in terms of the Pauli matrices $\mathbf{\hat{\sigma}} =(\hat{\sigma}^x,  \hat{\sigma}^y, \hat{\sigma}^z)$  as 
\begin{align}\label{eq:phase-point}
\hat{A}_\alpha  = \hat{\wp} ( \mathbf{r}_\alpha ), \qquad    \hat{\wp} ( \mathbf{r} ) \equiv (\hat{\sigma}^0  + \mathbf{r} \cdot  \mathbf{\hat{\sigma}} )/2, 
\end{align}
with the vectors $\mathbf{r}_{ (0,0) } = (1, 1, 1) $,  $\mathbf{r}_{ (0,1) } = (-1, -1, 1) $, $\mathbf{r}_{ (1,0) } = (1, -1, -1) $, and $\mathbf{r}_{ (1,1) } = (-1, 1, -1)$ \cite{Wootters1987}.  
Note that we have also included a $\hat{\sigma}^0$ term to allow sampling from unnormalized density matrices. For a system with $N$ spin-1/2 the phase space spans by $4^N$ points, i.e., $\alpha= \{\alpha_1, \alpha_2, ... \alpha_N \}$. 
The time evolution evolves under the classical dynamics of phase-space variables as
\begin{align}\label{eq:dtwa}
\langle  \hat{O} \rangle (t) = \sum_{\alpha} w_\alpha(0) \mathcal{O}_\alpha^W(t)  \approx \sum_{\alpha} w_\alpha(0) \mathcal{O}_\alpha^{W, \mcl } (t),
\end{align}
where $\mathcal{O}_\alpha^W$ is the Weyl symbol for the operator $\hat{O} $ and $ \mathcal{O}_\alpha^{W, \mcl } (t)$  represents 
the classical evolution.  $w_\alpha(0) $ is the initial Wigner function on the discrete many-body phase space. 
It factorizes for every spin $i$, i.e., $w_\alpha(0) = \prod_{i=1}^N  w_{\alpha_i}^{[i]} $, where the superscript $[i]$ denotes the phase space for spin $i$. 
Similarly, for the initial density matrix we have $\hat{\rho}(0) = \prod_{i=1}^N  \hat{\rho}^{[i]} $.
For the initial state with spins pointing in the $-z$ direction,  $w_{\alpha_i}^{[i]} = \mTr[ \hat{\rho}^{[i]}(\hat{z}) \hat{A}_{\alpha_i} ]/2$ yields  $w_{(0,0)}^{[i]} = w_{(0,1)}^{[i]} =0 $ and  $w_{(1,0)}^{[i]} = w_{(1,1)}^{[i]} =1/2$ for every spin $i$. 
This is illustrated in Fig. \ref{Fig:dtwa}(a), where the three sets of lines (two horizontal, two vertical, and two diagonal) correspond to the probability of a measurement outcome. 
This means the probability for a spin being in the $+z$ and $-z$ direction is $0\%$ and $100\%$, respectively. 
Similarly, the probabilities for a spin being in the $\pm x$ and $\pm y$ directions are  $50\%$ and $50\%$, respectively.

To solve the open-system dynamics we use the quantum master equation in Lindblad form 
\begin{align}\label{eq:qme}
\frac{d}{dt}\hat{\rho} = - i [\hat{H}, \hat{\rho}] + \sum_{i} \Bigl(  \hat{c}_i \hat{\rho} \hat{c}_i^\dagger - \frac{1}{2} \{ \hat{c}_i^\dagger \hat{c}_i, \hat{\rho} \}     \Bigr),
\end{align}
where the Hamiltonian $\hat{H}$ describes the coherent evolution and
the jump operators $\hat{c}_i$ correspond to the incoherent part of
the dynamics. While our OSDTWA approach is completely generic, we will
exemplify our method for a dissipative variant of the Ising model in a
transverse field \cite{Lee2011}, which is one of the most important
models in the analysis of open quantum many-body systems. The interest
in this model does not only stem from the paradigmatic character
similar to the transverse-field Ising model for closed
quantum systems \cite{Sachdev1999}, but also from its importance to
understand experimental results obtained in strongly interacting
Rydberg gases \cite{Carr2013,Malossi2014}. Its Hamiltonian has the conventional form 
$\hat{H}= (g/2) \sum_i \hat{\sigma}_i^{x} + (V/4) \sum_{ \langle i j \rangle } \hat{\sigma}_i^{z} \hat{\sigma}_j^{z}$,
where $g$ is the transverse field and $V$ is the nearest-neighbor
interaction. Dissipation is introduced via spin-flip operators $\hat{c}_i =
\sqrt{\gamma_d} \hat{\sigma}_i^-$, with $\gamma_d$ being the decay
rate of the up spins and $\hat{\sigma}_i^- = (\hat{\sigma}_i^{x} - i
\hat{\sigma}_i^{y} )/2$.  This model can
be realized using laser-driven Rydberg atoms, for which the spin-down
state corresponds to the atomic ground state and the spin-up state
refers to an excited Rydberg state.  Transitions between the states
are driven by a coherent laser with a Rabi frequency $\Omega =g$ and
the interaction $V$ describes a repulsive van der Waals interaction
$C_6/a^6$ determined by a $C_6$ coefficient at the lattice spacing $a$
\cite{Weimer2015a}.

%
%\begin{align}\label{eq:model}
%\hat{H}= \frac{g}{2} \sum_i \hat{\sigma}_i^{x} + \frac{V}{4} \sum_{ \langle i j \rangle } \hat{\sigma}_i^{z} \hat{\sigma}_j^{z}, 
%\end{align}
%

In the following, we obtain the dynamics of the interacting many-body
system by replacing the time evolution via classical trajectories as
described in Eq. \ref{eq:dtwa}.  We use classical spin variables
$S_i^\beta$, with $\beta=(x,y,z,0)$.  The initial states are sampled
on the discrete phase space according to the distributions encoding
the spin pointing down for all particles, i.e., we fix $S_i^z = -1$
and the spin components in the orthogonal direction are chosen
randomly as $S_i^x$, $S_i^y = \pm 1$ with equal probability. In
contrast to the closed DTWA \cite{Schachenmayer2015}, we also include
classical variables $S_i^0$, which encodes the local norm of a given
site and is initialized to $S_i^0=1$. 
This additional degree of
freedom is necessary because already the closed DTWA conserves the
norm of the Bloch vector only after averaging over all trajectories,
while our quantum-jump approach requires knowledge of the norm on the
level of a single trajectory.
Each spin of the state propagates under
the effective non-Hermitian Hamiltonian $ \hat{H}_i - i \gamma_d
\hat{\sigma}_i^+ \hat{\sigma}_i^-/2$.  The corresponding semiclassical
equations of motion are \cite{Note1}
\begin{align}
\dot{S}_i^x  &= - \frac{V}{2} S_i^y \sum_j S_j^z - \frac{\gamma_d}{2} S_i^x , \label{eq:Sx} \\ 
\dot{S}_i^y  &=  \frac{V}{2} S_i^x \sum_j S_j^z  - g S_i^z - \frac{\gamma_d}{2} S_i^y , \label{eq:Sy}\\
\dot{S}_i^z  &=  g S_i^y - \frac{\gamma_d}{2} (S_i^z + S_i^0), \\ 
\dot{S}_i^0  &=  - \frac{\gamma_d}{2} (S_i^z + S_i^0),
\end{align}
with the sum over $j$ being performed over the nearest neighbors of
the spin $i$. Here, the interaction terms are incorporated on
the level of a mean-field decoupling, as it is the case in the closed
DTWA. 
Importantly, this mean-field decoupling is performed on the level of a single trajectory, therefore the ensemble average does not correspond to the mean-field equations of motion for the density operator. If desired, it is also possible to include higher orders of the Bogoliubov-Born-Green-Kirkwood-Yvon (BBGKY) hierarchy of correlation functions in the phase-point operators \cite{Czischek2018}.
We numerically integrate the equations of motion using a
fourth-order Runge Kutta method.  The global norm $S^0 (t)= \prod_i^N
S_i^0 (t)$ decreases under the time evolution from its initial value
$S^0(0) = 1$. Once the global norm drops below a random number $r$
drawn from a standard uniform  distribution, a quantum jump occurs. Importantly,
this approach allows to use a high-order numerical integrator for both
the coherent and dissipative parts of the time evolution and thus
yields a higher order of accuracy compared to direct approaches to
solve the quantum master equation \cite{Daley2014, Johansson2012}. The
precise time $\tau$ of the quantum jump is determined by solving the
equation $S^0 (\tau) = r$.

Having determined the time of the quantum jump, we still need to
choose which of the jump operators (i.e., on which site) is actually
occuring. For this, we calculate the jump probability for spin $i$ by
$\delta p_i = \bigl( \prod_{j \neq i}^N S_j^0 \bigr) \times
\gamma_d(S_i^0 + S_i^z)/2$ \cite{Note1}.  The jump operator that is fired is then
chosen to occur at site $n$ such that $n$ is the smallest integer
satisfying $\sum_i^n P_i(\tau) \geq r$, where $P_i = \delta p_i/(
\sum_i^N \delta p_i)$ is the normalized spin probability
\cite{Johansson2012}. For the fired spin $n$, we set $S_n^z=-1$ and
choose $S_n^x$ and $S_n^y$ randomly as $\pm 1$ again with equal
probability.  For all other spins we normalize the spin fields by
$S_i^0$ as $S_i^\beta = S_i^\beta/S_i^0$.  We continue the time
evolution by generating a different $r$ and by repeating the above
procedure, see Fig. \ref{Fig:dtwa}(b). 
To avoid rare events leading to a divergence of the spin
variables, we clip each individual spin variable to
$|S_i^z|<\sqrt{3}$, which is the largest possible value that can be reached in an individual trajectory in a closed system. However, we find that this clipping is only necessary in the absence of interactions.
In Fig. \ref{Fig:dtwa}(b) we also show $\langle S^z (t)
\rangle$, which initially displays oscillatory behavior and then
eventually reaches a steady state.

%We show the time evolution of $S^z(t)$ for a single spin and $g/\gamma_d =5$.  
%By averaging over many initial configurations we obtain the ensemble averaged observable $\langle S^z
%\rangle = (1/n_t)\sum_i^{n_t} S_i^z$, where $n_t$ is the number of trajectories.

\begin{figure}[t]
\includegraphics[width=1.0\linewidth]{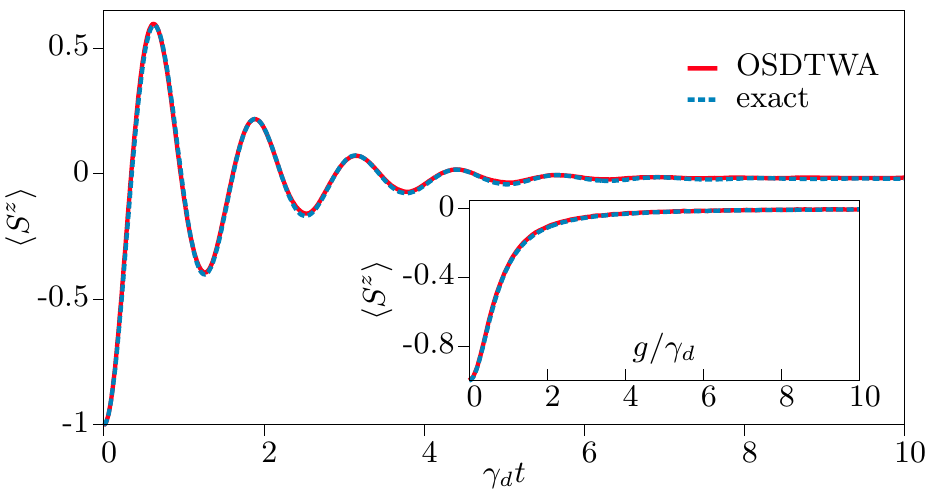}
\caption{Benchmarking against exact results for a single spin.
The time evolution of $\langle S^z (t) \rangle$, same as in Fig. \ref{Fig:dtwa}(b), is compared with the exact result $S^z_{\mexact} (t)$ for $g/\gamma_d = 5$.  
The inset shows the numerically obtained steady state from $\langle S^z (t) \rangle$ in the long-time limit and the exact steady state for varying $g/\gamma_d$. 
%The inset shows the residual of $S^z_{\mexact} (t_0) - \langle S^z (t_0) \rangle$  for step size $\Delta t$ and fixed $t_0 \gamma_d =50$, where the simulation result is averaged over $2$ million trajectories and $g/\gamma_d =5$.
%The dashed line is the fit proportional to $\Delta t^{6.15 \pm 0.22}$. 
%The horizontal line indicates the noise floor from the total number of trajectories.
 }
\label{Fig:exact}
\end{figure}

{\it{Benchmarking the OSDTWA.}---} In the following, we compare the OSDTWA to the time evolution of a
single spin, as in this case, the method does not introduce any
additional errors from the mean-field decoupling in
Eqs.~(\ref{eq:Sx}--\ref{eq:Sy})  and the sampling of the phase space in terms of a complete set of single-site operators is exact.  
We refer to this as the non-interacting case as it does not contain any spin-spin interactions.
Hence, the OSDTWA should match the
exact solution of the quantum master equation \cite{Breuer2002} in the
limit of vanishing step size of the numerical integration. In Fig. \ref{Fig:exact} we compare the simulation result of $\langle S^z (t) \rangle$ with the exact result $S^z_{\mexact} (t)$ for $g/\gamma_d =5$. 
Their comparison shows an excellent agreement; see also \cite{Note1}.
For the steady state $S^z_{\mexact} (t)$ yields the result $ S^z_{\mexact} = -1/(1+ 2\gt^2)$, with $\tilde{g}= g/\gamma_d$.
We therefore determine the numerical result of the steady state from $\langle S^z (t) \rangle$ in the long-time limit $t \gamma_d= 100$. 
In the inset of Fig. \ref{Fig:exact} we present the numerical and the exact result of the steady state as a function of $g/\gamma_d$. 
The steady state of the OSDTWA again agrees excellently with the exact steady state. Furthermore, we find the error in $\langle S_z\rangle$ scaling like $\Delta t^{6.15\pm 0.22}$ with the integration step size $\Delta t$ \cite{Note1}.

%For any numerical integration method, it is crucial to determine the
%scaling of the error with the size of the integration step $\Delta
%t$. To investigate this behavior, we examine the difference to the
%exact solution in the steady state as a function of $\Delta t$. 
%As shown in the inset of Fig.~\ref{Fig:exact}(b), we observe a power-law
%scaling of the error proportional to $\Delta t^{6.15\pm 0.22}$, which
%is even better than the $\Delta t^5$ scaling of the Runge-Kutta method
%for a single integration step. 
%This improvement can be attributed to
%the robustness of the steady state to step size errors
%\cite{Press1992}, which is a generic feature of open quantum systems.

Another important consequence of our particular choice of the
incorporation of quantum jumps is that the method becomes also exact
when the dynamics is governed by classical rate equations. In this
case, our approach yields a quantum-jump version of conventional
kinetic Monte-Carlo methods \cite{Bortz1975}.

\begin{figure}[t]
\includegraphics[width=1.0\linewidth]{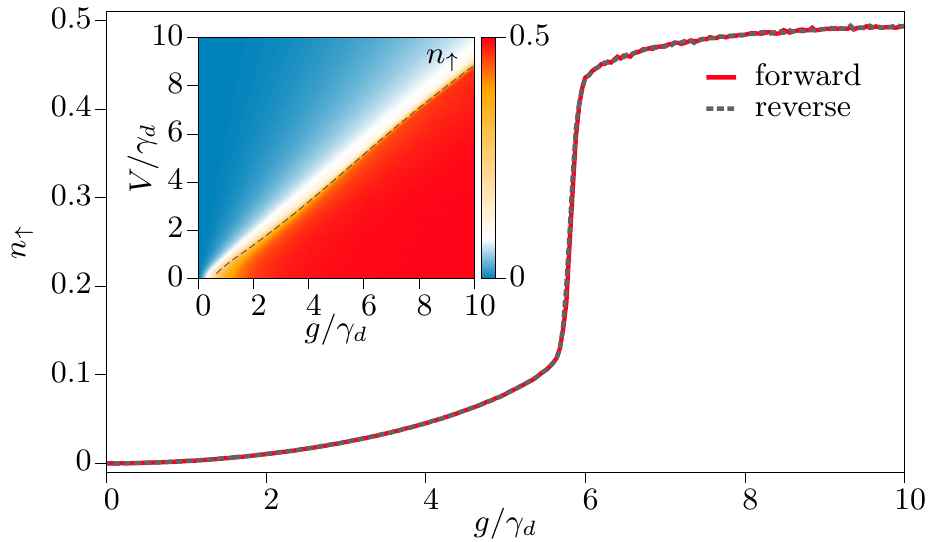}
\caption{Liquid-gas transition.
The spin-up density $n_\uparrow$ as a function of $g/\gamma_d$ for $V/\gamma_d = 5$ for both the forward (continuous line) and reverse sweep (dashed line). The inset depicts $n_\uparrow$ as a function of  $g/\gamma_d$ and $V/\gamma_d$, where the dashed line is the location of the susceptibility peak, see the main text. All results are shown for a $10\times 10$ lattice and $3200$ trajectories. 
 }
\label{Fig:sweep}
\end{figure}

 \begin{figure*}[]
\includegraphics[width=1.0\linewidth]{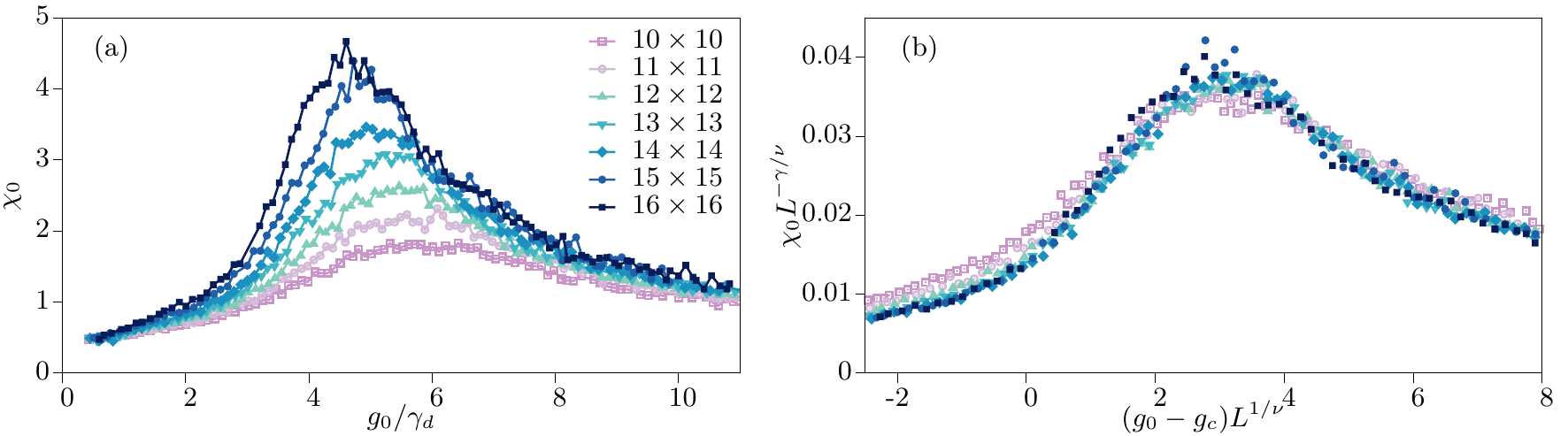}
\caption{Driven-dissipative criticality.  (a) Value $\chi_0$ and position $g_0$ of the susceptibility peak for varying system sizes between $10 \times 10$ and $16\times16$, derived from Gaussian fits to the susceptibility $\chi (g)=(\partial n_\uparrow / \partial g )$. Results were obtained using up to $12,000$ trajectories. (b) Universal scaling close to the critical point obtained by fitting the susceptibility data to the finite-size scaling function Eq.~(\ref{eq:scaling}), which yields the critical exponents $\gamma =1.69  \pm 0.07$ and  $\nu =0.99  \pm 0.04$,  and  the critical point $g_c/\gamma_d =2.94  \pm 0.14$.
 }
\label{Fig:sus}
\end{figure*}

{\it{Driven-dissipative criticality.}---} 
Let us now turn to the dissipative Ising model including the Ising interaction on a two-dimensional square
lattice. From variational calculations \cite{Weimer2015},
field-theoretical arguments \cite{Maghrebi2016}, tensor network
simulations \cite{Kshetrimayum2017}, and cluster mean-field theory
\cite{Jin2018}, it is known that the model exhibits a first-order
transition for sufficiently strong interactions $V$, when varying the
strength of the transverse field $g$. This transition can be
understood as a liquid-gas transition of spin-up particles and the
first-order transition line vanishes eventually in a critical point
when decreasing $V$ \cite{Marcuzzi2014,Weimer2015}. 
Importantly, this transition is not due to spontaneous breaking of the $\mathbb{Z}_2$ symmetry of the Hamiltonian (as this is already broken by the dissipation), but it is governed by the appearance of an emergent symmetry, similar to the liquid-gas transition in thermal equilibrium. 
Using mean-field
analysis, the critical point has been predicted to belong to the Ising
universality class \cite{Marcuzzi2014}, but it has not been possible
to analyze the critical behavior going beyond a mean-field treatment.

To demonstrate that the OSDTWA is capable of capturing fluctuations
beyond mean-field theory, we first consider a $10\times 10$ lattice
with periodic boundary conditions and $V/\gamma_d =5$.  We calculate
the spin-up density $n_\uparrow= ( 1 + \langle S^z \rangle )/2$ using
the steady state value of $\langle S^z (t) \rangle$ in a long-time
limit. Starting from the solution at $g=0$, we follow the steady state
for $g$ in the range $g/\gamma_d= [0, 10]$ using both a forward and a
reverse sweep of $g$.  In Fig. \ref{Fig:sweep} we show the results of
$n_\uparrow$ as a function of $g/\gamma_d$ for both cases of forward
and reverse sweeps. The perfect overlap demonstrates that the steady
state obtained within the OSDTWA is unique and not plagued by the
mean-field artifact of bistability \cite{Lee2011,Marcuzzi2014}.
%
%\new{despite the fact that the OSDTWA also performs a decoupling of the
%interaction on a mean-field level}. 
%
In addition, the results of
$n_\uparrow$ manifest a first-order phase transition, since
$n_\uparrow$ undergoes a steep jump around $g/\gamma_d =5.8$, which is
also in very good quantitative agreement with previous numerical
predictions \cite{Weimer2015}.  In the inset of Fig. \ref{Fig:sweep}
we show $n_\uparrow$ as a function of $g/\gamma_d$ and $V/\gamma_d$.
For intermediate and large $V/\gamma_d$, $n_\uparrow$ indicates a
sharp increase as $g/\gamma_d$ is increased across the first-order
transition. For small $V$, the change of $n_\uparrow$ appears in a much
broader region, suggesting that the first-order line
eventually terminates in a critical point.

To investigate the critical behavior of the model, we determine the
susceptibility $\chi (g)=(\partial n_\uparrow / \partial g )$ by
taking a numerical derivative of $n_\uparrow $ with respect to $g$.
We fit $\chi (g)$ to the Gaussian function $f(g) = \chi_0 \exp\bigl(
(g-g_0)^2 /(2 \sigma^2) \bigr)$, with $\chi_0 $, $g_0$, and $\sigma$
being the fitting parameters.  $g_0$ gives the location of the
susceptibility peak, which is indicated as a dashed line in the inset
of Fig. \ref{Fig:sweep}.  $\chi_0 $ is the height of the
susceptibility peak, which we use to determine the critical point
below.

To identify the critical point and its properties, we calculate
$\chi_0(g_0)$ for varying system sizes between $10 \times 10 $ and $16
\times 16 $ sites. All simulations employ periodic boundary
conditions. In Fig.~\ref{Fig:sus}(a), we show $\chi_0(g_0)$ for the
different system sizes, which displays a susceptibility peak diverging
with system size. The precise nature of this divergence is controlled
by the critical exponents of the transition, which in the framework of
finite-size scaling theory \cite{Cardy1996} can be captured as
\begin{align}\label{eq:scaling}
\chi_0 (g_0, L) = L^{\gamma/\nu} f\bigl( (g_0-g_c)L^{1/\nu} \bigr) ,
\end{align}
where $L$ is the linear dimension of the system, $g_c$ is the critical
point, and $\gamma$ and $\nu$ are the critical exponents. Due to the
hyperscaling relations \cite{Huang1987}, which can also be expected to
hold for steady-state transitions obeying thermal statistics, two
critical exponents are sufficient to fix all others as well. The
analytic scaling function $f(x)$ is then expanded as a fourth-order
polynomial and fitted to the results of $\chi_0$, which allows us to
determine the critical parameters in the thermodynamic limit.  From
the fit, we obtain $g_c/\gamma_d = 2.94 \pm 0.14$, $\gamma =1.69 \pm
0.07$, and $\nu=0.99 \pm 0.04$. Using these results, we observe all
susceptibility data to collapse on a single line, see
Fig.~\ref{Fig:sus}(b), which demonstrates that we have correctly
identfied the critical exponents. Remarkably, the values of $\gamma$
and $\nu$ are in very good agreement with $\gamma=7/4$ and $\nu=1$ of
the 2D classical Ising model, i.e., the dissipative Ising model
belongs to the same universality class. Furthermore, the OSDTWA value
for the critical point $g_c/\gamma_d =2.94 \pm 0.14$ lies between the
predictions from the variational principle ($g_c/\gamma_d = 2.28$
\cite{Weimer2015}) and cluster mean-field theory ($g_c/\gamma_d =
4.88$ \cite{Jin2018}).

Strikingly, the OSTDWA is able to capture fluctuations beyond
mean-field theory, although the Ising interaction is decoupled on a
mean-field level. This can be attributed to the fact that classical
fluctuations are correctly accounted for in our quantum-jump approach,
while quantum fluctuations are irrelevant at the transition due to the
presence of a dynamical symmetry yielding an effective field theory at
finite temperature \cite{Sieberer2013}. Interestingly, in this
approach, the quantum fields are gapped and can be mapped onto
classical fluctuation fields by means of a Hubbard-Stratonovich
transformation \cite{Sieberer2016}, which is conceptually very similar
to the random choices of the $S^{x,y}$ fields following a quantum jump
within the OSDTWA.

{\it{Conclusions and outlook.}---} We have presented a novel
simulation approach for an open quantum system based on the discrete
truncated Wigner approximation. For the paradigmatic dissipative Ising
model on a square lattice, we arrive at the first prediction of its
critical behavior beyond mean-field theory, which we find to be
consistent with the 2D Ising universality class. Importantly, our
method can be expected to give reliable results for a large class of
open quantum many-body systems governed by a dynamical
symmetry. Additionally, despite its computational simplicity, our
OSTDWA method can be used to obtain novel insights into non-critical
many-body problems that are notoriously hard to simulate, such as
strongly interacting Rydberg polaritons
\cite{Gorshkov2011a,Peyronel2012,Pistorius2020}. Finally, in future
studies it will be interesting to see whether the OSTDWA can also
capture open many-body systems displaying non-thermal critical
behavior, as it has been recently reported for quantum versions of
absorbing state models \cite{Carollo2019}.

\textit{Note added:} During preparation of our manuscript, we became aware
of a related work employing a quantum state diffusion approach to the
discrete TWA \cite{Huber2022}.

%% Acknowledgments---%%
%
{\it{Acknowledgments.}---}  This work was funded by the Volkswagen Foundation, by the Deutsche Forschungsgemeinschaft (DFG, German Research
Foundation) within Project-ID 274200144 -- SFB 1227 (DQ-mat, Project No. A04), SPP 1929 (GiRyd), and under Germany’s Excellence Strategy--EXC-2123 QuantumFrontiers--390837967.

%\textcolor{red}{Include new SM refs in the citation to the SM, by adding ``, which includes {\textbackslash}cite$\{$...$\}$''}

\bibliographystyle{myaps}
\bibliography{../bib}

\end{document}

\title{Supplemental Material for ``Driven-dissipative criticality within the discrete truncated Wigner approximation''}
\author{Vijay Pal Singh}
\affiliation{Institut f\"ur Theoretische Physik, Leibniz Universit\"at Hannover, Appelstra{\ss}e 2, 30167 Hannover, Germany}
\affiliation{Zentrum f\"ur Optische Quantentechnologien and Institut f\"ur Laserphysik, Universit\"at Hamburg, 22761 Hamburg, Germany}
\author{Hendrik Weimer}
\affiliation{Institut f\"ur Theoretische Physik, Leibniz Universit\"at Hannover, Appelstra{\ss}e 2, 30167 Hannover, Germany}
%
%
%\date{\today}
%
%%%------------------------------------------ ABSTRACT ----------------------------------------------------------------------------------
%
\maketitle
%
%---------------------------------------------------------------Introduction: BEGIN -------------------------------------------------------
%

\section*{Monte-Carlo wave-function formalism}\label{sec:MCWF}
We describe the open-system dynamics using the quantum master equation in Lindblad form 
\begin{align}\label{eq:qme}
\frac{d}{dt} \rho = - i [ H, \rho] + \sum_{i} \Bigl(  c_i \rho c_i^\dagger - \frac{1}{2} \{ c_i^\dagger c_i,  \rho \}     \Bigr),
\end{align}
where $\rho$ is the density matrix describing the ensemble average of a quantum system, $H$ is the system Hamiltonian, and $c_i$ are the collapse operators. 
We rewrite Eq.~(\ref{eq:qme}) as
\begin{align} \label{eq:qqme}
\dot{\rho } = - i (H_\meff \rho - \rho  H_\meff^\dagger) + \sum_{i}  c_i \rho c_i^\dagger, 
\end{align}
where 
\begin{align}\label{eq:Heff}
H_\meff  = H -  \frac{i}{2} \sum_i c_i^\dagger   c_i 
\end{align}
is the non-Hermitian effective Hamiltonian for the dissipative system. The term $ \sum_{i}  c_i \rho c_i^\dagger$ in Eq. \ref{eq:qqme} is referred to as the recycling term \cite{Daley2014}.
For a two-level system with decay from an excited state to the ground state, the non-Hermitian part can be thought of as removing amplitude from the excited state and the recycling term as restoring it in the ground state.  

Following Refs. \cite{Dalibard1992, Dum1992}, the time evolution of $\rho(t)$ can be described in terms of quantum jumps and nonunitary evolution between these jumps via \cite{Dum1992_2}
\begin{align}
\rho(t) = \sum_{n=0}^{\infty} \sum_{\gamma_1, ..., \gamma_n} \int_{t_0}^{t} dt_n \int_{t_0}^{t_n} dt_{n-1} ...  \int_{t_0}^{t_2}dt_1 S_{tt_n} c_{\gamma_n} S_{t_n t_{n-1} } ...  \, S_{t_2 t_1} c_{\gamma_1} S_{t_1 t_0}  \rho(t_0),
\end{align}
with
\begin{align}
S_{tt_0} \rho(t_0) = e^{- iH_\meff (t-t_0) }   \rho(t_0) e^{i H_\meff^\dagger (t-t_0) }.
\end{align}
The indices $n=0,1,2..$ and $\gamma_1,..., \gamma_n$ account for contributions to the density matrix from the subensemble that has undergone exactly $n$ quantum jumps at times $t > t_n \geq ... \geq t_1 \geq t_0$ with a sequence of realizations $\gamma_1,..., \gamma_n$. To construct a wave-function representation we use $\rho(t_0) = \sum_\alpha p_\alpha | \alpha \rangle \langle \alpha|$ ($0 \leq p_\alpha  \leq 1$ and $\sum_\alpha p_\alpha=1$) and obtain
\begin{align}\label{eq:mcwf}
\rho(t) =\sum_\alpha \sum_{n=0}^{\infty} \sum_{\gamma_1, ..., \gamma_n} \int_{t_0}^{t} dt_n \int_{t_0}^{t_n} dt_{n-1} ...  \int_{t_0}^{t_2}dt_1  |\phi(t|t_n, \gamma_n; ...; t_1, \gamma_1; \alpha) \rangle   \langle  \phi(t|t_n, \gamma_n; ...; t_1, \gamma_1; \alpha)| p_\alpha, 
\end{align}
where $|\phi(t|t_n, \gamma_n; ...; \alpha) \rangle$ are a hierarchy of system wave functions that follow 
\begin{align}
i \frac{d}{dt} | \phi(t|t_n, \gamma_n; ...) \rangle = H_\meff  |\phi(t|t_n, \gamma_n; ...)\rangle 
\end{align}
for $t \geq t_n$ with initial condition $|\phi(t_0|\alpha) \rangle = |\alpha \rangle$.

   Eq.~(\ref{eq:mcwf}) enables a wave function simulation of the density matrix in terms of Monte Carlo wave functions (MCWFs). 
The initial state is sampled from the density matrix $\rho(t=0)$ at time $t=0$. 
The simulation procedure is as follows. 
\begin{enumerate}
\item We start with a normalized wave function $| \psi (0)  \rangle$.

\item We choose a random number $r$ between $0$ and $1$, which corresponds to the probability that a quantum  jump occurs. 

\item We numerically propagate $| \psi (t)  \rangle$ according to $H_\meff $, i.e., 
\begin{align}
i \frac{d}{dt} | \psi (t)  \rangle =H_\meff | \psi (t)  \rangle 
\end{align}
and determine the quantum jump time $t$ by solving $\langle \psi (t) | \psi (t)  \rangle = r$, where $\langle \psi (t) | \psi (t)  \rangle  $ is the norm. 

\item \label{step3} At time $t$, the resulting jump operator projects the system into one of the renormalized states given by 
\begin{align}
| \psi (t^{+})  \rangle = \frac{c_i |\psi (t^{-})  \rangle }{  \langle  \psi (t^{-})| c_i^\dagger  c_i  |\psi (t^{-})  \rangle^{1/2}  },
\end{align}
where $|\psi (t^{-})  \rangle $ is the wave function propagating in time up to time $t$ and $|\psi (t^{+})  \rangle $ is the state after the jump.  
The corresponding jump operator is chosen such that $l$ is the smallest integer satisfying $\sum_{i=1}^l P_i(t) \geq r$, where $P_i  = \delta p_i/\sum_{i}  \delta p_i$ is the normalized probability, with 
\begin{align} \label{eq:prob}
\delta p_i =  \langle  \psi (t)| c_i^\dagger c_i |\psi (t)  \rangle.
\end{align}

\item We now continue the time evolution with the renormalized state from step \ref{step3} and draw a new random number $r$. 
This process is repeated until the final simulation time is reached. 

\end{enumerate}
The MCWF approach  determines the time evolution on the level of a single trajectory. By repeating the time evolution over different initial wave functions, an ensemble of the trajectories is obtained, giving the time evolution that is equivalent to the quantum master equation in Eq.~(\ref{eq:qme}), i.e.,  
\begin{align} \label{eq:equivalance}
\rho(t) = \biggl\langle \biggl\langle   \frac{| \psi(t) \rangle  \langle  \psi (t)| }{ \langle  \psi (t)| \psi (t)  \rangle} \biggr \rangle \biggr \rangle,
\end{align}
where $\langle \langle \quad \rangle \rangle  $ denotes an average over many trajectories.

\section*{Open-system discrete truncated Wigner approximation (OSDTWA) }
\subsection*{Derivation of equations of motion}
To obtain a semiclassical description of open quantum systems, we establish an open-system variant of the discrete truncated Wigner approximation. 
This includes a semiclassical approximation of the density matrix, i.e., we sample the initial state from the discrete truncated Wigner approximation (DTWA). Furthermore, the time evolution is obtained on the level of individual trajectories within the quantum-jump formalism, which is discussed in the previous section. 
As an application of our method, we consider the dissipative Ising model on a square lattice. 
The transverse-field Ising Hamiltonian is 
\begin{align}\label{eq:model}
H= \frac{g}{2} \sum_i \sigma_i^{x} + \frac{V}{4} \sum_{ \langle i j \rangle }  \sigma_i^{z} \sigma_j^{z}, 
\end{align}
where $g$ is the transverse field and $V$ is the nearest neighbor interaction. $\sigma_i^\alpha$ are the Pauli operators at site $i$, with $\alpha=(x,y,z)$.
The dissipative terms in Eq.~(\ref{eq:qme}) occur in terms of spin-flip operators $c_i = \sqrt{\gamma_d} \sigma_i^-$, with $\gamma_d$ being the decay rate of the up spins and $\sigma_i^{\pm} = (\sigma_i^{x} \pm i \sigma_i^{y} )/2$. 
Eq.~(\ref{eq:qme}) results in
\begin{align} \label{eq:qqme2}
\dot{\rho } &= - i ( H_i \rho - \rho H_i) - \frac{\gamma_d}{2}  \sg_i^{+} \sg_i^{-} \rho - \frac{\gamma_d}{2} \rho \sg_i^{+} \sg_i^{-}   + \gamma _d \sg_i^{-} \rho \sg_i^{+}  \\
& =  - i (H_{\meff,i} \rho - \rho  H_{\meff,i}^\dagger) + \gamma_d  \sg_i^{-} \rho \sg_i^{+}   \label{eq:rhoHeff} ,    
\end{align}
where 
\begin{align}\label{eq:Heff}
H_{\meff,i}  = H_i - i \frac{\gamma_d}{2}  \sg_i^{+}  \sg_i^{-}.
\end{align}
In the quantum-jump approach the system propagates under $H_\meff$ between two quantum jumps, hence the phase-space variables also evolve under  $H_\meff$ within the OSDTWA. The generalized phase-point operator is given by 
\begin{align}\label{eq:phase-point}
\wp_i ( \mathbf{r}_i ) = (\sigma_i^0  + \mathbf{r}_i \cdot  \mathbf{\sigma}_i )/2, 
\end{align}
where $\mathbf{\sigma}_i =(\sigma_i^x,  \sigma_i^y, \sigma_i^z)$  is the Pauli vector, and $\sigma_i^0$ is included to allow sampling from unnormalized density matrices. In our semiclassical approximation we use classical spin variables $S_i^\beta$, with $\beta=(x, y, z, 0)$. 
With the parametrization $\mathbf{r}_i(t) = (S^x_i, S^y_i, S^z_i)$, Eq.~(\ref{eq:phase-point}) results in 
\begin{align}
\tilde{\wp}_i( \mathbf{r}_i )= \frac{1}{2} (S_i^0  \cdot  \mathbb{1}_i +  \mathbf{r}_i  \cdot \mathbf{\sigma}_i ) = 
\begin{pmatrix}
\frac{1}{2}( S_i^0  + S_i^z )  &   S_i^{-}   \\ 
S_i^{+}   & \frac{1}{2}( S_i^0 - S_i^z  ) 
\end{pmatrix}.
\end{align}
$S_i^0$ represents the norm of the spin state, i.e., $S_i^0= \mTr[\tilde{\wp}_i]$. 
The time evolution of $\tilde{\wp}_i(t)$ under $H_\meff$ as described in Eq.  \ref{eq:rhoHeff} is given by 
\begin{align}
\begin{pmatrix}
  \frac{1}{2}( \dot{S}_i^0  + \dot{S}_i^z )    &    \dot{S}_i^{-}  \\ 
 \dot{S}_i^{+}  &    \frac{1}{2}( \dot{S}_i^0 -  \dot{S}_i^z)    
\end{pmatrix} 
=  \frac{1}{2}
\begin{pmatrix}
g S_i^y - \gamma_d ( S_i^0+ S_i^z)  &   i g S_i^z -  S_i^{-} ( \gamma_d + 4 i V_j)  
\\ 
 -i g S_i^z -  S_i^{+} ( \gamma_d - 4 i V_j)      &  -g S_i^y 
\end{pmatrix},
\end{align}
which results in
\begin{align}
\dot{S}_i^x  &= - 2 V_j S_i^y - \frac{\gamma_d}{2} S_i^x , \label{eq:Sx} \\ 
\dot{S}_i^y  &=  2 V_j S_i^x   - g S_i^z - \frac{\gamma_d}{2} S_i^y , \label{eq:Sy}\\
\dot{S}_i^z  &=  g S_i^y - \frac{\gamma_d}{2} (S_i^z + S_i^0), \label{eq:Sz} \\ 
\dot{S}_i^0  &=  - \frac{\gamma_d}{2} (S_i^z + S_i^0),  \label{eq:S0}
\end{align}
where $V_j \equiv (V/4) \sum_j S_j^z$ and the index $j$ runs over the nearest neighbors of site $i$.

The quantum jump probability, as described in Eq.~(\ref{eq:prob}), is determined from the decay of the state, which is 
\begin{align}\label{eq:prob1}
\delta p_i = \mTr[ \tilde{\wp}_i c_i^\dagger c_i ] = \gamma_d \mTr[ \tilde{\wp}_i  \sigma_i^{+} \sigma_i^{-}] = \frac{\gamma_d}{2} (S_i^0 + S_i^z).
\end{align}
For more than a single site, we calculate the jump probability at site $i$ via 
\begin{align}\label{eq:prob2}
\delta p_i = \frac{\gamma_d}{2}  (S_i^0 + S_i^z)  \times \prod_{k \neq i}^N S_k^0,
\end{align}
where $\prod_{k \neq i}^N S_k^0$ is a normalization factor coming from all other sites than the site $i=k$.

\subsection*{Implementation of the OSDTWA}
As we describe in the main text,  we sample the initial state from the discrete truncated Wigner approximation. 
This is done by fixing  $S_i^z=-1$ and randomly choosing $S_i^x, S_i^y =\pm 1$  with equal probability.
$S_i^0$ represents the local norm of a given site and is initialized to $S_i^0=1$. 
Using a fourth-order Runge Kutta method, we numerically integrate the equations of motion in Eqs.~(\ref{eq:Sx}--\ref{eq:S0}). 
Once the global norm $S^0(t)= \prod_{i=1}^N S_i^0 (t)$ drops below a random number $r$ drawn from a standard uniform distribution, a quantum jump occurs. The precise quantum jump time $\tau$ is determined by solving the equation $S^0(\tau)=r$ using the Ridders' method. 
To choose the  jump operator, we calculate the jump probability for spin $i$ using Eq.~(\ref{eq:prob2}). The jump operator that is fired is then chosen to occur at site $n$ such that $n$ is the smallest integer satisfying $\sum_{i=1}^n P_i(\tau) \geq r$, where $P_i = \delta p_i/(\sum_i^N \delta p_i)$ is the normalized spin probability. For the fired spin $n$, we set $S_n^z=-1$ and choose $S_n^x$ and $S_n^y$ randomly as $\pm 1$ again with equal probability. 
For all other spins we normalize the spin fields by $S_i^0$ as $S_i^\beta = S_i^\beta/S_i^0$. 
Afterwards, we continue the time evolution by generating a different $r$ and by repeating the above procedure.
Note that these steps are equivalent to the MCWF formalism discussed in the previous section. 
With the above procedure, we obtain the time evolution of a single trajectory, which undergoes several quantum jumps and nonunitary evolution between these jumps. 
By generating different initial configurations and different times of the quantum jumps, we obtain an ensemble of the trajectories. 
As the key observable, we calculate the averaged spin value in the $z$ direction, i.e., $\langle  S^z \rangle =(1/n_t) \sum_i^{n_t} S_i^z$, 
where $n_t$ is the number of trajectories.

\begin{figure}[t]
\includegraphics[width=1.0\linewidth]{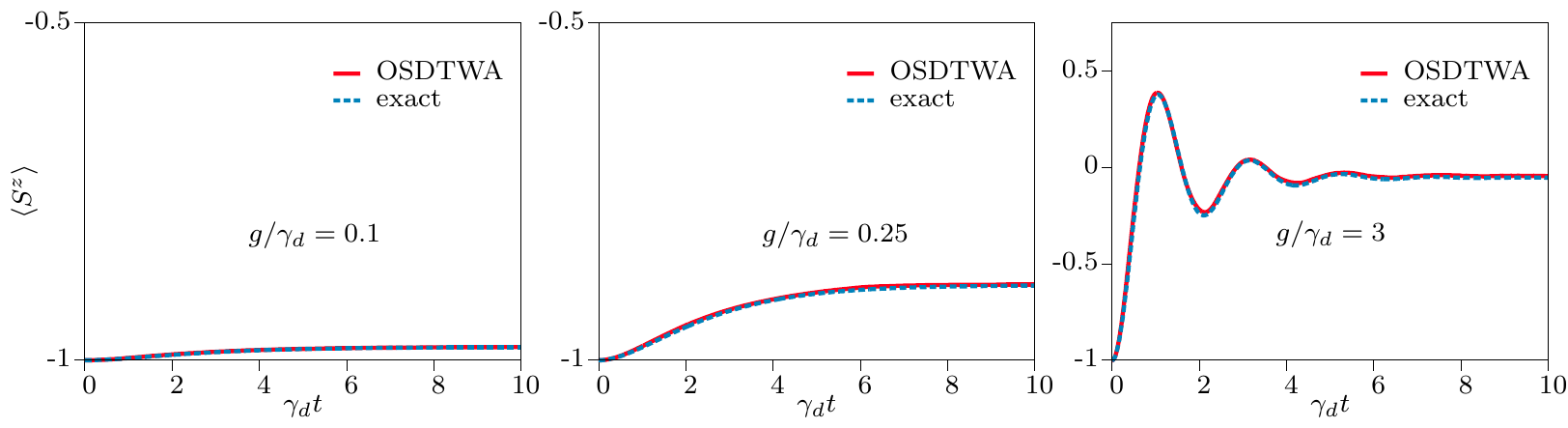}
\caption{Comparison with exact results.
The time evolution of $\langle S^z (t) \rangle$, averaged over $10^5$ trajectories, is compared with the exact result $S^z_{\mexact} (t)$ for $g/\gamma_d=0.1$, $0.25$, and $3$.
 }
\label{Fig:S1}
\end{figure}

To confirm that the OSDTWA is exact in the noninteracting limit, we compare the OSDTWA to the time evolution of a single spin, as in this case, the method does not include any additional errors from the interaction terms in Eqs.~(\ref{eq:Sx}) and  (\ref{eq:Sy}). Hence, the OSDTWA should match the exact solution of the quantum master equation in the limit of vanishing step size of the numerical integration. 
For the exact result, we have \cite{Breuer2002}
\begin{align} \label{Eq:exact}
 S^z_{\mexact} (t) =  \frac{2 \gt^2 }{1+ 2 \gt^2} \Biggl(1 - \exp(-3 \tt/4) \Bigl(  \cos( \sqrt{\gt^2-1/16} \tt )   +    \frac{3}{4 }  \frac{ \sin( \sqrt{\gt^2-1/16} \tt ) }{ \sqrt{\gt^2-1/16} }   \Bigr)      \Biggr) - 1, 
\end{align}
with $\tilde{g}= g/\gamma_d$ and $\tt = t \gamma_d$.
In Fig. \ref{Fig:S1} we compare the time evolution of $\langle S^z (t) \rangle$ with the exact result $S^z_{\mexact} (t)$ for $g/\gamma_d =0.1$, $0.25$, and $3$. 
The OSDTWA and the exact result show excellent agreement. 
The weakly-damped limit occurs when $S^z_{\mexact} (t)$ yields an oscillatory motion in time, 
which is visible for $g/\gamma_d=3$ in Fig. \ref{Fig:S1}. 
The oscillatory behavior vanishes when $\tilde{g}$ approaches the critical value $ \tilde{g}_\mcritical =0.25$ and 
Eq.~(\ref{Eq:exact}) becomes $ S^z_{\mexact} (t) = \bigl[1-\exp(-3\tt/4) \times (1+3\tt/4) \bigr]/9 -1$. 
We show this exact result and the OSDTWA for $g/\gamma_d=0.25$  in Fig. \ref{Fig:S1}.
For  $\tilde{g}$ lower than $\tilde{g}_\mcritical$, the system is overdamped, as we show the result at $g/\gamma_d=0.1$  in Fig. \ref{Fig:S1}. 

\begin{figure}[ ]
\includegraphics[width=0.50\linewidth]{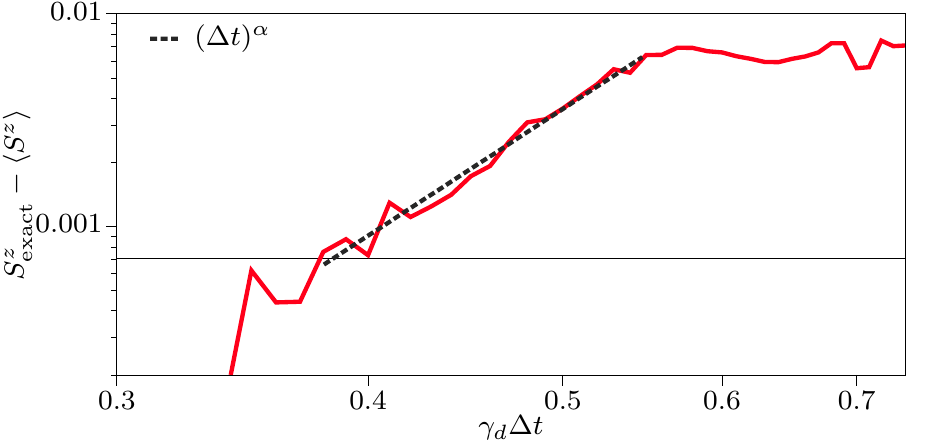}
\caption{The residual of $S^z_{\mexact} (t_0) - \langle S^z (t_0) \rangle$  for step size $\Delta t$ and fixed $t_0 \gamma_d =50$, where the simulation result is averaged over $2$ million trajectories and $g/\gamma_d =5$.
The dashed line is the fit proportional to $\Delta t^{6.15 \pm 0.22}$. 
The horizontal line indicates the noise floor from the total number of trajectories.
 }
\label{Fig:S2}
\end{figure}

For any numerical integration method, it is crucial to determine the
scaling of the error with the size of the integration step $\Delta
t$. To investigate this behavior, we examine the difference to the
exact solution in the steady state as a function of $\Delta t$. 
As shown in Fig.~\ref{Fig:S2}, we observe a power-law
scaling of the error proportional to $\Delta t^{6.15\pm 0.22}$, which
is even better than the $\Delta t^5$ scaling of the Runge-Kutta method
for a single integration step. 
This improvement can be attributed to
the robustness of the steady state to step size errors
\cite{Press1992}, which is a generic feature of open quantum systems.

\subsection*{Offdiagonal coherence at the transition}
\begin{figure}[ ]
\includegraphics[width=0.50\linewidth]{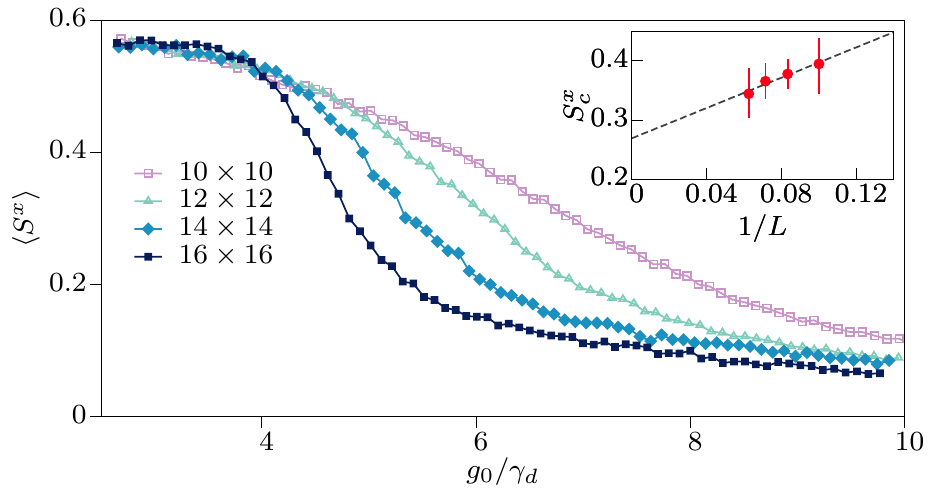}
\caption{Steady state value of $ \langle S^x \rangle$ as a function of $g_0$ for varying system sizes between $10\times 10$ and $16\times 16$, 
where $g_0$ represents the location of line of the liquid-gas transition. Results are obtained using trajectory numbers between $640$ and $1920$. 
We determine the values of $ \langle S^x \rangle$ at the critical point $g_c(L)$ and show them as a function of $1/L$ in the inset plot, with $L$ being the linear dimension of the system. 
The error bar is due to the uncertainty in the location of $g_c(L)$; see text. 
Using a linear fit (continuous line), we determine the critical value $S^x_c = 0.27 \pm 0.01$ in the thermodynamic limit. 
 }
\label{Fig:Sx}
\end{figure}

One important question concerning phase transitions in open quantum
many-body systems is their degree of quantumness, especially since
they inherently involve mixed quantum states. To address this
question, we analyze the steady-state value of the $ \langle S^x
\rangle$ observable, which is an offdiagonal coherence with respect to
the $ \langle S^z \rangle$ observable capturing the liquid-gas
transition of the dissipative Ising model. We focus on $ \langle S^x
\rangle$ at the location $g_0$ of the susceptibility peak (i.e., along
the curve of the liquid-gas transition), whose value and location are
used to characterize the transition in the main text.  In
Fig. \ref{Fig:Sx} we show $ \langle S^x \rangle$ as a function of
$g_0$ for various system sizes between $10\times 10$ and $16\times
16$.  For $g_0$ below $g_0/\gamma_d < 4$, the value of $\langle S^x
\rangle$ is large and depends very weakly on the system size.  In the
intermediate regime of $g_0/\gamma_d$ between $4$ and $8$, $\langle
S^x \rangle$ undergoes a noticeable change depending on the system
size. This suggests that off-diagonal coherences are important to
describe the critical behavior.  We determine the location $g_c(L)$ of
the critical point from the susceptibility curves shown in the main
text, where $L$ is the linear dimension of the system.  We calculate
$\langle S^x_c \rangle$ at $g_c(L)$, finding a $1/L$ decay with system
size, see the inset of Fig. \ref{Fig:Sx}. From this, we can determine
the critical value in the thermodynamic limit as $S^x_c (L \rightarrow
\infty) = 0.27 \pm 0.01$. This shows that offdiagonal coherence is
persisting at criticality.

\section*{Critical behavior of the dissipative XYZ model}

In the following, we use the OSDTWA to examine the criticality of the
dissipative XYZ model on a square lattice, which exhibits a phase
transition of the steady state between a paramagnetic and a
ferromagnetic phase, as first suggested by mean-field predictions of
Ref.~\cite{Lee2013}.  This dissipative model has been analyzed using
tensor network \cite{Rota2017,Kshetrimayum2017} and cluster mean-field
\cite{Jin2016} methods. Using a tree tensor network approach, it was suggested that the phase transition belongs to the universality class of the two-dimensional (2D) Ising model \cite{Rota2017}, however, the
finite-size scaling analysis was limited to system sizes up to $6
\times 6$.  Here, we use our OSDTWA to simulate the dynamics of the
dissipative XYZ model on a square lattice for system sizes between
$8\times 8$ and $14\times 14$, as we describe below.

We consider spin-$1/2$ particles that interact via anisotropic
Heisenberg interactions, whose Hamiltonian has the form
\begin{align}\label{eq:xyz}
H= \sum_{ \langle i j \rangle } \bigl(  J_x  \sigma_i^{x} \sigma_j^{x}  + J_y  \sigma_i^{y} \sigma_j^{y}  + J_z  \sigma_i^{z} \sigma_j^{z}   \bigr),
\end{align}
where $J_{\alpha}$ are the spin-spin couplings between nearest-neighbors, with $\alpha=x, y, z$.
Dissipation in the system is realized via spin-flip operators $c_i = \sqrt{\gamma_d} \sigma_i^-$, 
which have the same form as in the dissipative Ising model considered in the main text. 
If $J_x$ and $J_y$ are anisotropic, i.e., $J_x \neq J_y$, the system can undergo a dissipative phase transition from a paramagnetic state with zero magnetization in the $xy$ plane to a ferromagnetic state with nonzero magnetization in the $xy$ plane \cite{Lee2013}. 

\begin{figure*}[b]
\includegraphics[width=1.0\linewidth]{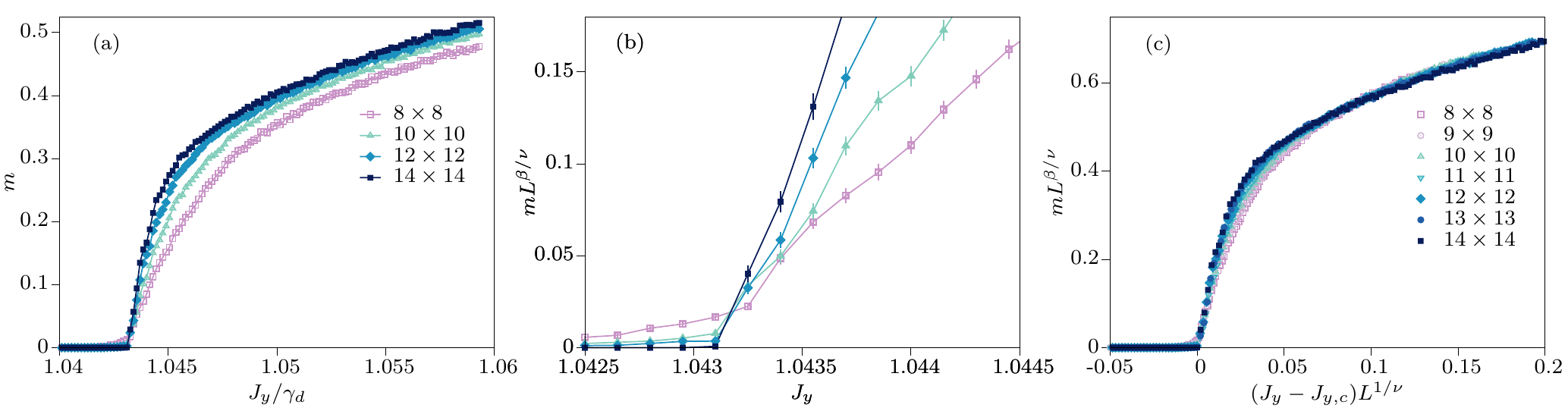}
\caption{Critical behavior of the dissipative XYZ model.  (a) Magnetization $m$ as a function of $J_y/\gamma_d$  for varying system sizes between $8 \times 8$ and $14\times14$. We use $J_x/\gamma_d=0.9$ and $J_z/\gamma_d=1$. 
Results were obtained using trajectory numbers between $600$ and $1,000$. 
(b) Scaled magnetization $mL^{\beta/\nu}$ with $\beta=1/8$ and $\nu=1$ according to the 2D Ising universality class. Using a linear interpolation between the data points, all curves cross around $J_y = J_{y,c} \approx 1.0431 \pm 0.0002$, with the error being given by the spacing between the data points. Error bars on the individual curves were obtained from the statistical sampling error; see text. 
(c) $m L^{\beta/\nu}$ versus $(J_y - J_{y,c})L^{1/\nu}$ shows a data collapse for all system sizes, confirming the transition belonging to the 2D Ising universality class.
 }
\label{Fig:S3}
\end{figure*}

We implement the OSDTWA following the description provided in the
previous section. In particular, we generate the initial states
according to the DTWA, i.e., we fix $S_i^z=-1$ and randomly choose
$S_i^x, S_i^y =\pm 1$ with equal probability.  For the time evolution
we evolve the initial state under the effective Hamiltonian based on
Eq.~(\ref{eq:rhoHeff}).  This results in the equations of motion
\begin{align}
\dot{S}_i^x  &=  - 2 S_i^y  \sum_j  J_z S_j^z   + 2 S_i^z  \sum_j  J_y S_j^y   -   \frac{\gamma_d}{2} S_i^x ,   \\ 
\dot{S}_i^y  &=    2 S_i^x  \sum_j  J_z S_j^z   - 2 S_i^z  \sum_j  J_x S_j^x  -   \frac{\gamma_d}{2} S_i^y ,   \\
\dot{S}_i^z  &=  - 2 S_i^x  \sum_j  J_y S_j^y   + 2 S_i^y  \sum_j  J_x S_j^x   - \frac{\gamma_d}{2} (S_i^z + S_i^0),  \\ 
\dot{S}_i^0  &=  - \frac{\gamma_d}{2} (S_i^z + S_i^0),   
\end{align}
where the index $j$ runs over nearest neighbors corresponding to the site $i$. 
Again , we numerically solve these equations of motion with a fourth-order Runge Kutta method. 
The time evolution undergoes random quantum jumps, which are implemented following the Monte-Carlo procedure described above.
To identify the transition, we calculate the steady-state magnetization defined as 
\begin{align}\label{eq:m}
m = \bigl \langle (m_x^2 + m_y^2)^{1/2} \bigr\rangle , \quad \text{with} \quad m_\alpha = \frac{1}{N} \sum_{i}^N S_i^\alpha, 
\end{align}
where $\langle ... \rangle$ represents the trajectory average and $N$ is the total number of lattice sites.  
For our analysis, we choose $J_x/\gamma_d=0.9$ and  $J_z/\gamma_d=1$, and vary  $J_y/\gamma_d$ to determine the transition. 
We use the integration time step $\gamma_d \Delta t=0.005$ and a maximum time of $\gamma_d t=3000$, 
for which the time evolution of $m(t)$ reaches a well-defined steady state. 
In Fig.~\ref{Fig:S3}(a), we show the OSDTWA steady-state value of the magnetization for various system sizes between $8\times8$ and $14 \times 14$. Increasing $J_y$ triggers a phase transition between a paramagnet having zero and a ferromagnet having nonzero magnetization, respectively.

To capture the precise nature of the transition we use the finite-size scaling behavior
\begin{align}\label{eq:scaling}
m(J_y,  L) = L^{-\beta/\nu} \bar{m} \bigl(   (J_y - J_{y,c})L^{1/\nu}   \bigr),
\end{align}
where $\beta$ and $\nu$ are the critical exponents, $L$ is the linear
dimension of the system, and $J_{y,c}$ is the location of the critical
point \cite{Cardy1996}.  We follow the scaling behavior based on
Eq.~(\ref{eq:scaling}) and therefore scale the magnitude as $m
L^{\beta/\nu}$, where we use the previously suggested values
$\beta=1/8$ and $v=1$ of the two-dimensional Ising model
\cite{Rota2017}, see Fig.~\ref{Fig:S3}(b).  The different system size
results cross at $J_y= J_{y,c} \approx 1.0431\pm 0.0002$, which we
identify as the critical point. Error bars can be estimated from
the standard deviation $\Delta m = \bigl( \langle m^2 \rangle - \langle
m \rangle^2 \bigr)^{1/2}$ , resulting in an error of $\Delta
m/\sqrt{N_t-1}$, where $N_t$ is the number of trajectories. We note
that this prediction for the critical point is in good agreement with
both the cluster mean-field analysis ($J_{y,c} = 1.03$) \cite{Jin2016}
and the tree tensor network approach ($J_{y,c} = 1.07\pm 0.02$)
\cite{Rota2017}.  In Fig.~\ref{Fig:S3}(c), we plot $m L^{\beta/\nu}$
versus $(J_y - J_{y,c})L^{1/\nu}$, which results in all data collapsing
onto a single curve, i.e., we have used the correct critical exponents
to describe the phase transition, confirming the transition belonging
to the universality class of the two-dimensional Ising model.

\bibliographystyle{myaps}
\bibliography{../bib}